\documentclass[12pt,preprint]{aastex}

\usepackage{graphicx}

\newcommand{\pol}{polo\"{\i}dal }
\newcommand{\tor}{toro\"{\i}dal }
\newcommand{\be}{\begin{equation}}
\newcommand{\ee}{\end{equation}}
\newcommand{\p}{\partial}
\newcommand{\nab}{{\bf \nabla}}

\newcommand{\BB}{\mbox{\bf B}}
\newcommand{\vv}{\mbox{\bf v}}

\slugcomment{Submitted to ApJ June 18th 2002, Accepted August 22nd 2002}

\shorttitle{Ideal MHD jet launching from resistive accretion disks}
\shortauthors{Casse \& Keppens}

\begin{document}

\title{Magnetized Accretion-Ejection Structures:\\
 2.5D MHD simulations of continuous Ideal Jet launching from\\ 
resistive accretion disks}

\author{Fabien Casse \& Rony Keppens}
\affil{ Institute for Plasma Physics Rijnhuizen, \\
P.O. Box 1207, 3430 BE Nieuwegein, The Netherlands \\
fcasse@rijnh.nl, keppens@rijnh.nl}

\begin{abstract}
We present numerical magnetohydrodynamic  (MHD) simulations of a magnetized
accretion disk launching trans-Alfv\'enic jets.  These simulations,
performed in a 2.5 dimensional time-dependent polytropic resistive MHD
framework,  model a resistive accretion disk threaded by an initial
vertical magnetic field. The resistivity is only important inside the disk,
and is prescribed as $\eta=\alpha_m V_AH\exp(-2Z^2/H^2)$ , where  $V_A$ stands
for Alfv\'en speed, $H$ is the disk scale height and the coefficient
$\alpha_m$ is smaller than unity.   By performing the 
simulations over several tens of dynamical disk timescales, we show that
the launching of a collimated outflow occurs self-consistently and  the
ejection of matter is continuous and quasi-stationary. These are the
first ever simulations of resistive accretion disks launching non-transient
ideal MHD jets. Roughly 15 \% of accreted mass
is persistently ejected.  This outflow is safely characterized as a jet
since the  flow becomes super-fastmagnetosonic, well-collimated and reaches
a quasi-stationary state. We present  a complete illustration and
explanation of the `accretion-ejection' mechanism that leads to jet
formation from a magnetized accretion disk. In particular, the magnetic
torque inside the disk brakes the matter azimuthally and allows for
accretion,  while it is responsible for an effective magneto-centrifugal
acceleration in the jet.  As such, the magnetic field channels the disk
angular momentum and powers the jet acceleration and collimation. The jet
originates from the inner disk region where equipartition between thermal
and magnetic forces is achieved. A hollow, super-fastmagnetosonic shell of
dense material is the natural outcome of the inwards advection of a
primordial field.
\end{abstract}


\section{Introduction}

\subsection{Observational data}\label{s:obs} 

Jets are ubiquitous phenomena in astrophysics since they are observed
around young stellar objects (YSOs), galactic compact objects (neutron
stars, X-ray binaries) and in some active galactic nuclei
\citep{Livi97}. These outflows are called jets because they display a very
good collimation far from the central object and reach high
velocities. Indeed, jets have terminal velocities of roughly $400 \,\, {\rm
km/s}$ around YSOs  (e.g. \citet{Garc01} and references therein), or reach
a fraction of the light speed for X-ray binaries \citep{Mira99} and
Fanaroff-Riley class I (FRI) jets. Relativistic motions are also detected
in FRII jets launched from active galactic nuclei \citep{Verm94}. Any model
to explain the creation of such outflows must provide a mass source for the
jets, achieve their collimation and contain a mechanism to accelerate mass
within the jet.

All astrophysical systems mentioned have accretion disks  associated with
them. Observed correlations between emission from the accretion disk and
from the jet provide evidence that the  jets are launched from the disks
directly.  In YSOs, since the work of \citet{Cabr90},  it is clear that the
disk luminosity is correlated with the light coming from optical forbidden
lines emitted in the jets (see also \citet{Hart95}). In galactic systems, a
similar correlation occurs, as for example in the first detected
microquasar GRS 1915+105 \citep{Mira98}. In these systems the disk
radiation is characterized by X-ray emission while an infrared and a
delayed radio emission seem to be the signature of a periodic ejection
phenomenon \citep{Eike98}.  As the energy emitted from the jets is a
synchrotron emission, the presence of a magnetic field has to be taken into
account for the ejection.  A recent study done by \citet{Serj98} has shown
that for AGNs, a link exists between an emission in the optical range
(believed to originate from the disk) and a radio-synchrotron emission
(associated with the jet).  Summarizing, observational evidence is mounting
that independent of the precise nature of the central accreting object (YSO
to AGN), magnetized jets are propelled from surrounding disks.

\subsection{Accretion-ejection models}

The most promising model for such `accretion-ejection' structures is based
on a scenario where a large scale magnetic field threads an accretion
disk. The presence of the magnetic field can be explained by the
advection of interstellar magnetic field \citep{Mous76} and/or by the
local production of a magnetic field thanks to an effective disk dynamo 
\citep{Reko00}. 
This model has been developed since the seminal work of \cite{Blan82},
using a magnetohydrodynamic (MHD) approach. It was shown that
the magnetic field can azimuthally brake the matter inside the disk
(carrying off angular momentum allowing accretion) 
and accelerate matter above the disk surface. The
collimation of the flow is achieved via magnetic tension due
to the presence of a \tor component of the magnetic field 
\citep{Love76,Blan76,Heyv89,Saut02}. The magnetic field provides an
effective alternative to the radially outward
transport of disk angular momentum by viscosity. The interaction of
the magnetic structure with the disk plasma can create a MHD Poynting flux
leaving the disk along the magnetic surface \citep{Ferr95}. This energy
flux can then be converted into kinetic energy of the matter within the
jet. Because the mass density in the jet is smaller than in the disk, it
is thereby possible to reach high terminal velocities for a given amount of
angular momentum removed from the disk. In this `accretion-ejection'
model, the mass of the jet is fed from the accretion disk. 
This mechanism requires the disk to be in an equilibrium where the
vertical thermal pressure gradient overcomes both the magnetic
pinching of the disk (if one considers a bipolar topology of the magnetic
field) and the gravitational compression. \citet{Ferr95} have shown 
that the only stable equilibrium configuration meeting that
requirement is an accretion disk where equipartition between thermal
and magnetic pressure prevails, in order to avoid both a too strong
magnetic compression and the magneto-rotational instability
(e.g. see the review by \citet{Balb98}). In this model, the outflow does 
not require an external mechanism to ensure a good collimation, since it
is achieved by the magnetic field itself. Moreover, the presence of a
central object only plays a role through its gravitational field alone
and there is
no specific interaction between its disk and its magnetospheric or its 
radiative environment presumed. Obviously, this scenario
can be applicable to every system owning a magnetized accretion disk, 
such as the systems mentioned in section~\ref{s:obs}.

\noindent The first observation of a jet has been done in 1918 by Curtis
for the optical jet seen around M87 \citep{Curt18}. Since this observation,
many other jets have been observed and monitored
over several decades without drastic
changes seen in their structures. Does this mean that a model to
describe Magnetized Accretion-Ejection Structures
(MAES) has to be completely stationary? The answer is not so easy to
give. We know that the ejection has been present over several decades in
most of the systems (except in X-ray binaries where the phenomenon seems to
be transient or periodic). This time can be compared with the dynamical
time of a MAES, as characterized by the rotation period of the
matter at the inner radius of the MAES. This dynamical time is  \be
\tau_{dyn} = \frac{2\pi}{\Omega_K} = 2\pi\sqrt{\frac{R^3_i}{GM_*}}
\label{intro1}
\ee
\noindent where $R_i$ is the inner radius of the MAES with a Keplerian
rotation profile around a central object of mass $M_*$. In YSOs, where
$M_*\sim 1M_{\odot}$ and $R_i\sim 0.1 AU$, the dynamical time is of the
order of ten days while in AGN, where $M_*\sim 10^8 M_{\odot}$ and $R_i\sim
10 R_S$ ($R_S$ being the Schwarzschild radius), the dynamical time is
typically two days. For the microquasar-type systems, the mass of the
stellar black hole is typically $10M_{\odot}$ and then the dynamical
time becomes 2 milliseconds with the same $R_i$. Since these times are
much shorter than the observed existence time of the associated jets, the
models producing jets have to be close enough to a steady-state to yield
permanent outflows over many dynamical time periods.

\noindent The approaches developed so far to explain persistent jet
launching can roughly be classified in two classes. One class uses a
semi-analytical formulation assuming stationarity while the other class
tries  to model at least a part of the structure using a time-dependent
numerical MHD code. The former contains all stationary self-similar studies
as well as the work by \citet{Ogil01} which uses Taylor expansions of
physical quantities  to study the dynamics of the launching region in
detail.  Some self-similar studies model both the accretion disk and the
super-Alfv\'enic jet in a stationary framework  using a variable separation
method with various levels of assumptions
\citep{Ward93,Ferr93a,Li95,Ferr95,Li96,Ferr97,Cass00a,Cass00b}. These
studies require some dissipative mechanisms (ambipolar diffusion or
magnetic resistivity) to occur inside the disk so that matter can cross the
magnetic surfaces and achieves an accretion motion. Although these
semi-analytical studies bring deep insight and reveal analytical relations
that apply generally to  the accretion-ejection mechanism,  the assumption
of self-similarity introduces geometric restrictions on the solutions.  The
other class of studies groups all work done using 2.5D MHD time-dependent
simulations (three dimensional assuming axisymmetry). Within this class,
two categories of study can be distinguished: one which aims at a numerical
calculation of the jet  alone \citep{Usty95,Ouye97,Kras99} while the second
one tries to model both the disk and the resulting outflow
\citep{Ushi85,Mats96,Kuwa00,Kato02}.  Stationary calculations of pure ideal
MHD jets necessarily treat the disk as a boundary condition. This makes it
difficult to determine whether  the prescribed quantities at the base of
the jet are in agreement with conditions prevailing in a magnetized
accretion disk.  Works that have tried to model  thick
\citep{Mats96,Kuwa00} or thin disks \citep{Kato02} launching outflows have
been done using either ideal MHD or uniformly resistive calculations
throughout the 
computational domain.  While these simulations demonstrate outflows from
the disk, the structure is very unstable and ejection only occurs during
very few dynamical times. This is in conflict with the observed stability
of jets mentioned above. The discrepancy in ideal MHD studies arises from
 modeling of the magnetized accretion disk subject to the frozen-in
condition, so that the magnetic field is continuously advected towards the central object. This gives rise to a central magnetic
field accumulation which will ultimately halt the accretion process
itself. Moreover, in order to be stationary, it would assume an infinite
magnetic reservoir feeding the process by some ad hoc outer boundary
conditions. On the other hand, calculations assuming uniform resistivity
throughout the computational domain have to justify the presence of such a
dissipative phenomenom outside the accretion disk. Our simulations will
require neither an infinite reservoir of magnetic field (if the magnetic
structure reaches a steady-state) nor resistivity outside the accretion disk.

\noindent The aim of the present paper is to be a first convergence point
between the two classes of works mentioned previously. Indeed, in this
paper we present 2.5D time-dependent MHD computations starting from an
initial configuration close to a self-similar one. Most importantly, we aim
to model continuous ideal MHD jet launching from a {\em resistive}
accretion disk threaded by a large scale magnetic field.  Treating the
accretion disk magnetic structure in a resistive manner is the key point to
achieving a robust ejection process. Instead of an unrealistic continuous
magnetic reservoir, this model only requires  the presence of a primordial
field. The magnetic field can attain a  quasi-stable configuration by
achieving  a balance between inwards advection, outward diffusion and
tensional retraction.  The organization of the paper is as follows: in the
next section, we present the MHD formalism used in this work and also the
initial conditions of our simulations. In the third section, we present the
numerical code we have used, and comment on the computational grid and the
boundary conditions. In the fourth section, we present our simulations and
give a complete description of the accretion-ejection mechanism. In the
last section we conclude and give an outlook to forthcoming work.

\section{Magnetized accretion disk}

We start with the time-dependent MHD equations governing the dynamics of
both the accretion disk and the outflow, in a non-relativistic
framework. In a second subsection, we provide all details on the initial
conditions of our simulations. Some necessary conditions have to be
fulfilled by this initial set up, in order to achieve a jet launching.  For
simplicity, we consider axisymmetric structures.  This assumption has a big
effect on the stability of the structure,  since it suppresses all
non-axisymmetric instabilities that may occur in a full three-dimensional
framework \citep{Kim00} .

\subsection{MHD equations}
The mass conservation equation, in an axisymmetric time-dependent description, 
can be written as  
\be
\frac{\p \rho}{\p t} + \nab\cdot(\rho \vv_p) = 0,
\label{mhd1}
\ee
\noindent where $\rho$ is the plasma density and $\vv_p$ is the
polo\"{\i}dal component of the velocity vector. The momentum conservation
takes into account three forces acting on the
plasma, namely thermal pressure gradient, the Lorentz force as well as the
gravitational force. The momentum $\rho\vv$ conservation reads
\be 
\frac{\p \rho\vv}{\p t} + \nab\cdot(\vv\rho\vv -\BB\BB) +
\nab(\frac{B^2}{2}+ P) + \rho\nab\Phi_G= 0,
\label{mhd2}
\ee
\noindent where $\BB$ is the magnetic field, $P$ the thermal pressure and
$\Phi_G=-GM_*/(R^2+Z^2)^{1/2}$ is the gravity potential created by the
central object. Coordinates $(R,Z)$ are cartesian coordinates in the \pol
plane.
The induction equation governs the evolution of the magnetic field
\be 
\frac{\p \BB}{\p t}= - \nab\cdot(\vv\BB - \BB\vv) - \nab\times(\eta {\bf J}),
\label{mhd3}
\ee
\noindent where the magnetic resistivity $\eta$ is meant to arise from 
turbulence occuring within the accretion disk. To enable
matter to cross stable (nearly steady-state) magnetic surfaces
in the disk, one has to consider transport phenomena within the accretion
disk. The current density ${\bf J}$ is directly 
related to the magnetic field by the Amp\`ere-Maxwell equation
\be
{\bf J} = \nab\times\BB.
\label{mhd4}
\ee
\noindent In the above equations units have been chosen where $\mu_o=1$.
In order to close the system, we need an energy
equation. 
In our study, we replace the energy equation by a very
simple polytropic relation
\be
P= K\rho^{\gamma},
\label{mhd5}
\ee
\noindent where the polytropic index $\gamma=C_p/C_v$ is the ratio of
specific heats and $K$ is a constant related to the sound speed 
by $C^2_S=K\gamma\rho^{\gamma-1}$.
In reality, the complexity of the turbulence occuring in an accretion disk
(microscopic reconnection, kinetic transport, etc...) may not be
adequately described by a fluid approach but may need treatment using a kinetic
theory framework. 

\subsection{Initial conditions}     

In what follows, we describe the initial
conditions for all physical quantities, namely the
density, the velocity and the magnetic field. The initial accretion disk
configuration is close to a self-similar configuration used in analytical
MAES studies. A modification was needed in order to take into account the
presence of the symmetry axis. The self-similar structure of the
flow decomposes every physical quantity as the product of a radial power-law
and a function of the polar angle.

i) \underline{Disk scale height.}
The first quantity to consider is the height of the accretion disk $H$. In
a flat accretion disk where $H$ is assumed constant, it will be difficult to
maintain a vertical equilibrium between thermal pressure and gravity, since
the latter decreases with radius. A more natural scaling for the accretion
disk height is an increasing $H$ with radial distance. 
We choose to prescribe a linear proportionality
as $H=\varepsilon R$. As shown by
\cite{Ferr93a}, this choice is consistent with 
polo\"{\i}dal magnetic field lines bending at the surface of a thin
disk in order to enable acceleration of outgoing matter. In
all our simulations we choose the disk aspect ratio $\varepsilon = 0.1$,
which is consistent with a thin accretion disk. 

ii) \underline{Mass density.}
The vertical profile of the density is a decreasing function of altitude
$Z$ so that the density typically decreases over one disk scale
height. The radial stratification of density, together with Keplerian
rotation, is constrained by the vertical disk equilibrium. This is
because the vertical
equilibrium of the disk imposes that the sound speed within the disk is
proportional to $\Omega_KH$, where $\Omega_K$ is the Keplerian angular
velocity \citep{Shak73}. The density profile reads
\be
\rho(R,Z,t=0) = \frac{R_o^{3/2}}{(R_o^2+R^2)^{3/4}}\left[
max\left(10^{-6},(1-\frac{(\gamma-1)Z^2}{2H^2})\right)\right]^{1/(\gamma -1)},
\label{init1}
\ee   
\noindent where $R_o$ is a constant set equal to $4$ in our runs. This
offset radius in the denominator makes the density regular up to $R=0$.
The ``max''
function ensures that the initial density does not reach unphysical
values.
The vertical variation ensures a hydrostatic equilibrium in a thin
polytropic disk. The radial
exponent $-3/2$ is imposed by the radial behaviour of the sound speed. As
indicated previously, the sound speed scales 
as $\Omega_KH\propto R^{-1/2}$.
For our polytropic relation we have
\be
C_s^2 = \gamma\frac{P}{\rho} = K\gamma\rho^{\gamma-1} \propto R^{-1},
\label{init1bis}
\ee  
\noindent which for $\gamma=5/3$ gives $\rho\propto R^{-3/2}$. 
Note that we normalize all velocities with respect to the factor
$\Omega_KH$ evaluated at the inner radius $R_i$. Normalizing
distances to this
inner radius of the disk $R=R_i=1$, this normalization entails that
$\sqrt{GM_*}=1/\varepsilon$.
The $K$ constant in the
polytropic relation is set to unity, making the sound speed $C_s$ of
order $\Omega_KH$. Finally, the dimensionless specification of the density
given by Eq.~\ref{init1} is done with respect to a fiducial density at
the origin. Actual dimensional specifications for specific systems like
YSOs or AGNs are retrieved by providing appropriate values for the mass of
the central object, the inner disk radius, and the density through the 
observed mass accretion rate of the system.

iii) \underline{Rotation profile.}
Our simulations deal with non-relativistic disk and jet
structures. The simplest configuration for the azimuthal velocity is then
Keplerian. In fact, we prescibe $V_{\theta}$ as
\be
V_{\theta}(R,Z,t=0) = \Omega R =
(1-\varepsilon^2)\frac{R_o^{1/2}}{\varepsilon(R_o^2+R^2)^{1/4}}\exp{\left(-2\frac{Z^2}{H^2}\right)},
\label{init2}
\ee 
\noindent so that we have sub-Keplerian rotation with a deviation
from Keplerian of order $\varepsilon^2$.
This is needed due to the presence of both radial thermal and magnetic
pressure gradients, of the same order.
This sub-Keplerian rotation ensures a radial equilibrium of
the disk. The factor $1/\varepsilon$ comes from the
velocity normalization: the ratio of the sound speed to the Keplerian speed is
proportional to $\varepsilon$ in a thin accretion disk where radiative
pressure is neglected \citep{Shak73,Fran86}.

iv) \underline{Polo\"{\i}dal velocity.}
The initial configuration of the \pol  flow is a pure accretion motion, i.e.
with a radially inward velocity. Since the angular and sound
speed are scaled as $R^{-1/2}$, we will use the same shape for the
horizontal $V_R$ and vertical $V_Z$ components
to attain a coherent disk equilibrium,
\begin{eqnarray}
V_R(R,Z,t=0)&=& -m_s\frac{R_o^{1/2}}{(R_o^2+R^2)^{1/4}}\exp{\left(-2\frac{Z^2}{H^2}\right)},\\
V_Z(R,Z,t=0) &=& V_R(R,Z,t=0)\frac{Z}{R} .
\label{init3}
\end{eqnarray}
\noindent The constant $m_s$ is a parameter smaller than unity, which then
ensures  an initial subsonic \pol inflow. This parameter, tuning the
amplitude of the radial velocity, cannot be very small in order to insure
the jet launching. Indeed, \citet{Blan82} have shown that the magnetic
acceleration of matter in an ideal MHD jet can only occur if the \pol
magnetic surfaces are bent with an angle larger than $30^\circ$ at the disk
surface. In our simulations, the disk surface marks the transition between
resistive  and ideal MHD regimes. So the accretion disk must evolve
dynamically to a configuration producing a radial magnetic field typically,
of the order of the vertical  one at the disk surface. Now the radial
component of the magnetic induction equation in the case of a thin
accretion disk ($|V_Z| \ll |V_R|$ and $|\p B_Z/\p R|\ll|\p B_R/\p Z|$)
supporting a stationary structure ($\p/\p t=0$) can be reduced to 
\be
\eta\frac{\p B_R}{\p Z} \simeq -  V_RB_Z > 0\ .
\label{extra2}
\ee
\noindent This equation clearly shows that 1) the only
stationary magnetic configuration allowing accretion ($V_R<0$) necessarily
involves non-vanishing magnetic resistivity for crossing field lines and 2) if the amplitude of $V_R$ is too small, the
Blandford \& Payne criterion will never be fulfilled since $\p B_R/\p Z$
will be very small. In order to have both an initial sub-sonic
accretion motion and an initial configuration favorable to jet
launching, we choose a value of $m_s$ smaller but close to unity,
typically $0.3$ in our simulations. Numerical experiments with values
$m_s\ll 1$ indeed failed to launch jets as expected from equation
(\ref{extra2}). On the other hand, large values of $m_s\gg 1$ led to
numerical results displaying strong magnetic pinching of the disk, unable
to give rise to a vertical mass flux feeding the jet.  

v) \underline{Magnetic field.}
The VAC code (see section~\ref{s-vac}) offers several ways to ensure 
the evolution of the magnetic field to be divergence
free~\citep{Toth00}. Nevertheless, it is necessary to start the numerical
integration with a magnetic field structure where $\nab\cdot\BB=0$. 
Because of the symmetry conditions on the equatorial plane of the disk and on
the rotation axis, the radial magnetic field must
vanish at these locations, namely $B_R(R=0,Z)=B_R(R,Z=0)=0$. Furthermore,
to produce jets,
the magnetic field pressure should be roughly of
the same order of magnitude than the thermal pressure at the equatorial
plane \citep{Ferr95}. The simplest configuration satisfying all these
conditions at $t=0$ is a radially stratified vertical magnetic field
\begin{eqnarray}  
B_Z &=& \frac{R_o^{5/2}}{(R_o^2+R^2)^{5/4}}\frac{1}{\sqrt{\beta}}, \\
B_R &=& B_{\theta} = 0,
\label{init4}
\end{eqnarray}
\noindent where $\beta$ is the plasma beta parameter measuring the ratio of the
thermal pressure to the magnetic pressure at $Z=0$. This parameter will
always be of order unity in our simulations. 

vi) \underline{Resistivity.}
The anomalous magnetic resistivity $\eta$
is believed to arise from turbulence triggered
within the accretion disk. Since we perform our simulations in an
axisymmetric fashion, this parametrisation of the resistivity is meant to
incorporate non-axisymmetric turbulent dynamics, associated with the action
of a disk dynamo  and/or with turbulence associated with non-axisymmetric
MHD instabilities that can affect equipartion accretion disks
\citep{apjl}. This transport mechanism enables the \pol flow of
the disk to pass through the magnetic surfaces, without enforcing
advection. We
adopt a modified \citet{Shak73} prescription for the magnetic resistivity,
namely
\be
\eta=\alpha_mV_A\left.\right|_{Z=0}H\exp{\left(-2\frac{Z^2}{H^2}\right)},
\label{init5}
\ee 
\noindent parametrized by $\alpha_m$. Note that this resistivity profile
essentially vanishes outside the disk, and that it varies in time as the
equatorial Alfv\'en speed $V_A=B/\sqrt{\rho}$ gets adjusted.  In previous
self-similar studies \citep{Cass00a}, the values for $\alpha_m$ were of
order unity. Such high values required typical wavelengths of the
turbulence produced by magnetic resistivity to be of the order of the disk
scale height. This is problematic if we expect the turbulence to be
triggered inside the disk. In the present fully numerical treatment, we
will therefore consider smaller, more realistic values of
$\alpha_m$. Another difference with previous self-similar treatments is the
fact that we consider an isotropic magnetic resistivity.   We recall that
\citet{Cass00a} have shown that in a self-similar accretion-ejection
structure a special turbulence configuration is needed to ensure that the
magnetic torque brakes the matter in the disk and accelerates it above the
disk surface.  This was cast in a specific relation involving all three
transport coefficients, viscosity  $\eta_v$ and the \tor and \pol magnetic
resistivity ($\eta_m'$ \& $\eta_m$), and both magnetic and viscous
$\nab\cdot\mathsf{T}$ torques. This relation is 
\be 
\frac{\nab\cdot\mathsf{T}}{({\bf
J}\times\BB.{e_{\theta}})}\propto
\frac{\varepsilon\eta_m'\eta_v}{3\eta_m^2}
\label{init6}
\ee
\noindent and shows that if one wants a thin accretion-ejection structure
to have comparable magnetic and viscous torques,  an anisotropy between
$\eta_m$ and $\eta_m'$ is required (see \citet{Cass00a} but also
\citet{Ogil01}). In our numerical simulations, we neglect the effect
of a viscous torque, so relation~\ref{init6} will be fulfilled if a
turbulence configuration such that $\eta_v\leq\eta_m$ and
$\eta_m=\eta'_m$ is achieved, since we consider a thin accretion disk
($\varepsilon\ll 1$). In this paper, we will assume the magnetic
resistivity to be isotropic $\eta_m=\eta_m'=\eta$. 

\section{Numerical scheme and boundary conditions}
\subsection{Numerical code: VAC}\label{s-vac}

All simulations reported here are done with the Versatile Advection Code
(VAC, see \citet{Toth96a} and {\tt http://www.phys.uu.nl/}$\sim${\tt
toth}).  We solve the set of resistive, polytropic MHD equations under the
assumption of a cylindrical symmetry. The initial conditions described
above are time advanced using the conservative, second order accurate Total
Variation Diminishing Lax-Friedrich \citep{Toth96b} scheme with minmod
limiting applied on the primitive variables. We use a dimensionally
unsplit, explicit predictor-corrector time marching.  To enforce the
solenoidal character of the magnetic field, we apply a projection scheme
prior to every time step \citep{Bruc80}.

\subsection{Grid and Boundary conditions}
\begin{figure}[t]
\plotone{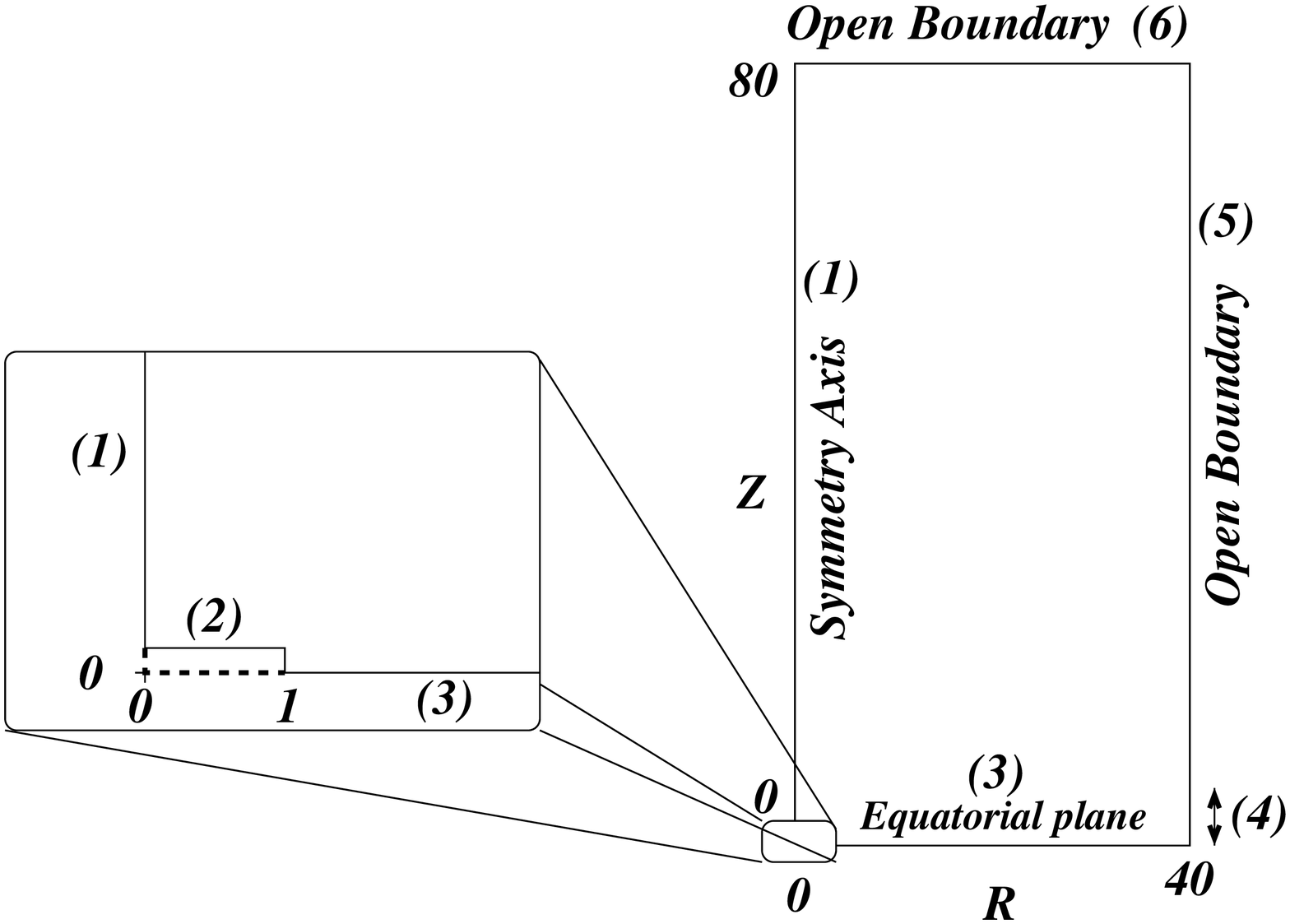}
\caption{Schematic representation of the computational domain. The bottom
left corner, containing the origin $(R,Z)=(0,0)$, is treated as
an internal boundary region in order
to avoid the gravitational singularity. This inner boundary enables us to
treat the inner radius of the accretion disk as a mass sink, where
the only requirement is to have $(V_R,V_Z)\leq 0$. This condition 
avoids unphysical mass flux entering the domain there. In the region
(4), we impose the accretion rate by prescribing $\rho V_p$, in order to
mimic the effect of the outer regions of the accretion disk. See the text
for details concerning each of the 6 boundary regions. \label{Figgrid}}
\end{figure} 

The purpose of our simulations is to model, under symmetry assumptions, a
magnetized accretion disk and the jets that can be launched from this
structure. Since bipolar jets are observed in the universe, we assume
that the system is symmetric with respect to the equatorial plane which
we label as $Z=0$. Moreover, we assume an axial
symmetry making $R=0$ a symmetry axis.
We use a rectangular grid spanning a physical
domain of $R=[0,40]$ and $Z=[0,80]$.
The grid resolution is
$154\times 304$ cells, with 2 ghost cells on each side for enforcing
boundary conditions. The cell widths and heights vary non-uniformly
throughout the physical domain, as both a radial and a vertical stretching
is applied. In effect, this achieves a higher resolution locally in the disk.

The problem setup involves
a Newtonian gravitational potential that has a singularity at the
origin. In order to avoid this problem without modifying the gravity
potential, we cut out several cells in
the bottom left corner of the grid
from the computational domain. 
This effectively introduces an internal
boundary region. In fact, we exploit 6 different boundary regions,
indicated in Fig.~\ref{Figgrid}.\\  
i) Inner boundary (Region (2) in Fig.~\ref{Figgrid})\\
\noindent The inner boundary is a rectangular area
excluded from 
the computational domain. It is taken to be
2 cells high in the $Z$-direction and 14 cells
wide in the $R$-direction. This number of cells corresponds to an inner
boundary which radially extends to $R=1$, the inner radius of our accretion
disk $R_i$. In this boundary region, we employ ``sink''
boundary conditions: in every ghost cell, the value of each quantity is
copied from the third cell row $Z_3$ just above the excluded domain. 
A restriction is imposed on the \pol velocity. Indeed, since we are
considering an accretion disk, we do not allow any positive (outward) mass
flux to exist in this region and we set the internal ghost cell values
for the \pol velocity as 
\begin{eqnarray}
V_R &=& min(V_R^{Z_3},0)\nonumber\\
V_Z &=& min(V_Z^{Z_3},0)
\label{bound1}
\end{eqnarray}
\noindent Under such a flow boundary condition, the matter that fills 
the zone $R<R_i$ and $Z>0$ can only originate from the disk itself
and not from the internal sink region. This is in contrast with studies done
so far trying to modelize similar structures \citep{Mats96,Kuwa00,Kato02},
where a modification of the gravitational potential is necessary. Such
modification would not be suitable to perform long-time integrations since
the accreted matter would rebound on the boundary after some time. \\
ii) Inflow region (Region (4))\\
\noindent We describe a magnetized accretion from 
an outer radius $R_e=40$ to its inner radius.
Obviously, real disks extend beyond 40 internal disk radii.
In order to mimic the effect of the unmodeled outer part of the
accretion disk, we impose the value of the \pol mass flux in the ghost
cells located at $R_e=40$ within one disk scale height only. 
This disk related part of the 
right boundary is therefore designed as a ``source'' 
region. The imposed accretion rate in this
region (4), is maintained at its initial value of the disk configuration by 
keeping $\rho V_R$ and $\rho V_Z$ fixed 
at this location. The remainder of the right boundary 
(region (5) in Fig.~\ref{Figgrid}) is treated as an open boundary
(zero gradient on all conserved quantities). \\
iii) Equatorial plane and polar axis (Regions (1) and (3))\\
\noindent These boundary areas are a combination of symmetric and antisymmetric
conditions. Symmetric and antisymmetric reflections of
the computed values within the computational domain correspondingly
set the ghost cell values.
See Table~\ref{Tab1} for the imposed conditions on the different physical
quantities.
\begin{table}[t]
\begin{tabular}{c|ccccccc}
 &$\rho$ & $V_R$ & $\Omega R$ & $V_Z$ & $B_R$ & $B_{\theta}$ & $B_Z$ \\ \hline
Equatorial plane & symm & symm & symm & asymm & asymm & asymm & symm\\
Symmetry axis & symm & asymm & asymm & symm & asymm & asymm & symm
\end{tabular}
\label{Tab1}
\caption{Boundary conditions for the physical quantities at the equatorial
plane ($Z=0$) and at the symmetry axis ($R=0$).}
\end{table}     

iv) Open boundaries (Region (5) and (6))\\
\noindent The last boundary regions we have to consider are the upper and
top right 
boundaries. These regions are expected to receive the expelled matter from
the disk, accelerated by the action of the magnetic field
combined with the disk rotation. Our approach to model non-reflecting outflow 
conditions is very simple:
the values of physical quantities in ghost cells are copied from the
nearest computational cell in the direction perpendicular to the boundary
surface. We check the possible influence of 
these boundary regions by changing the box size and repeating the simulations. 

\section{Accretion-ejection simulations}
In this section, we present the results of our simulations. We first
discuss the time evolution towards a quasi-equilibrium
state. We then analyse the accretion-ejection structure which forms
self-consistently and is maintained during
this simulation. The parameter set is resistivity level
$\alpha_m=0.1$, Mach sonic number $m_s=0.3$ and the magnetic pressure level
$\beta=1.67$. 
\begin{figure}[t]
\plotone{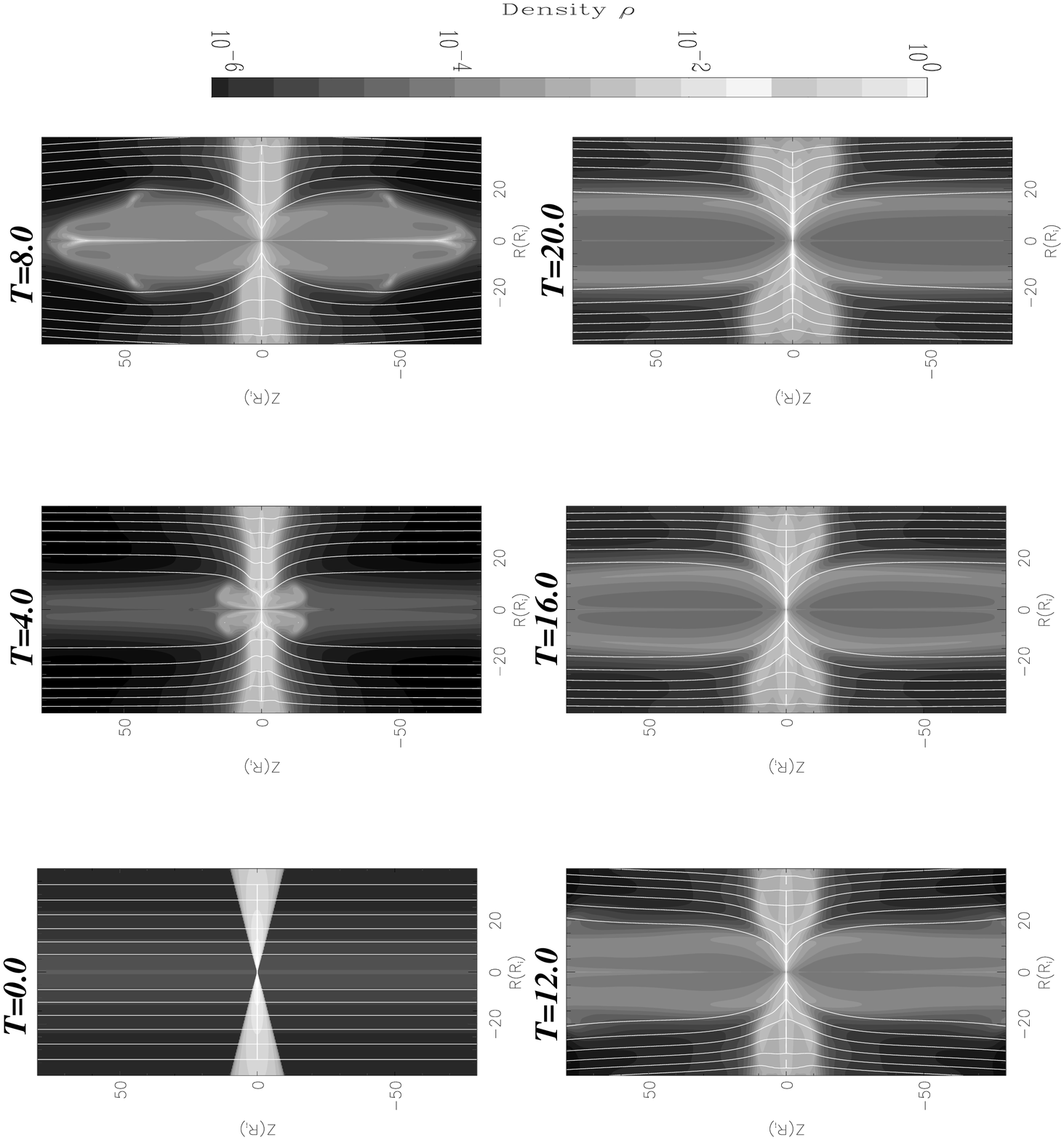}
\caption{The dynamical evolution of a thin
accretion disk threaded by an initial radially decreasing vertical magnetic
field, visualized in density levels (in greyscale) and \pol magnetic field 
lines. One clearly sees a denser region being ejected from the disk along
the symmetry axis, which can be identified with the jet. Collimation
is already present after few dynamical time scales.\label{Snap}}
\end{figure}
\subsection{Towards steady-state} 
The time evolution of the structure is displayed in Fig.~\ref{Snap}. The
time unit is $1/\varepsilon\Omega_K$ which in this simulation represents
$(2\pi\varepsilon)^{-1}=1.59$ rotation periods of matter at the inner
radius. We continued this simulation until $t=25$, covering 40
periods. During this timescale, we clearly see in Fig.~\ref{Snap} the jet
launching as a relatively dense outflow coming from the accretion disk,
more precisely confined to the inner region of the accretion disk for
radial distances $R\leq 20$.  The collimation of this outflow is obtained
within the computational domain.
\begin{figure}[t]
\plotone{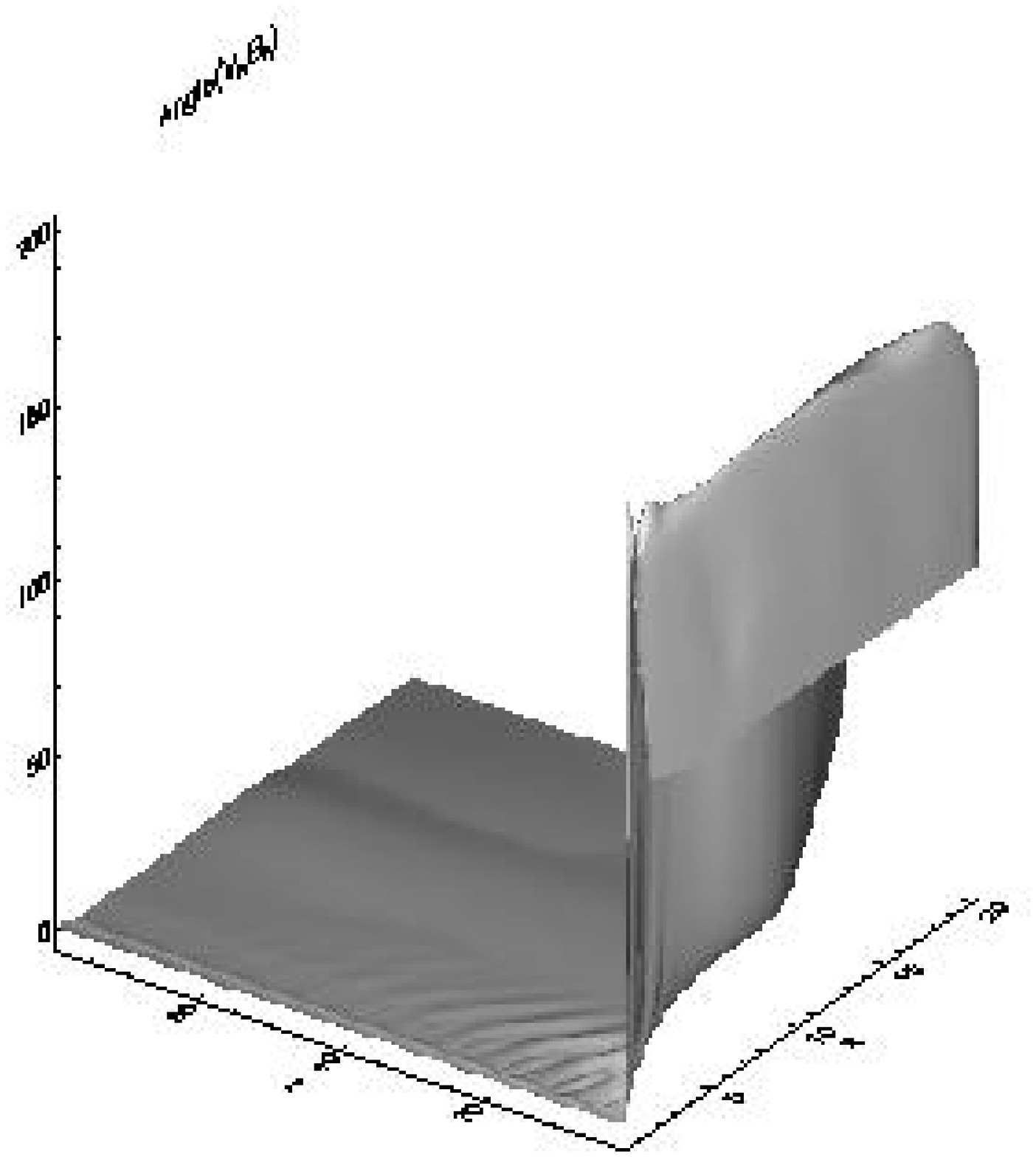}
\caption{Measure of the angle between the \pol velocity and the \pol
magnetic field within the accretion disk-jet region ($R \leq 20$) at
$t=19$. In the accretion disk, 
the two vectors are perpendicular because the magnetic field is almost
vertical and the accretion motion is radial. Moving upwards 
in the accretion disk,
the field is bent away from the symmetry axis and a growth of $B_R$ 
is occuring. Hence, the angle increases. Above the disk,
where the resistivity is vanishing, 
the two vectors tend to become parallel. In the
jet region ($R\leq 20$ and $Z\geq 20$), the angle between $\vv_p$ and $\BB_p$
is everywhere smaller than $3.5^{\circ}$, so that we effectively
obtain a near-perfect ideal MHD stationary jet.\label{angV}}
\end{figure}
To check whether the simulation is evolving towards a stationary state, 
we look at a quantity that would be equal to zero in a perfect ideal MHD
stationary state.
From the ideal MHD induction equation in an axisymmetric, stationary 
framework, one can deduce that
the \pol velocity and the \pol magnetic surfaces have to be
parallel. In Fig.~\ref{angV}, we display the angle between the two
vectors in the launching area of the jet at $t=19$. It can be seen
that above the accretion disk, this angle never
exceeds $3.5^{\circ}$. This near-perfect alignment is present from times
$t>17$ and is characteristic of a structure tending to a stationary
configuration. Nevertheless, even at this stage, 
the structure is still slowly evolving in time. 
\begin{figure}
\plotone{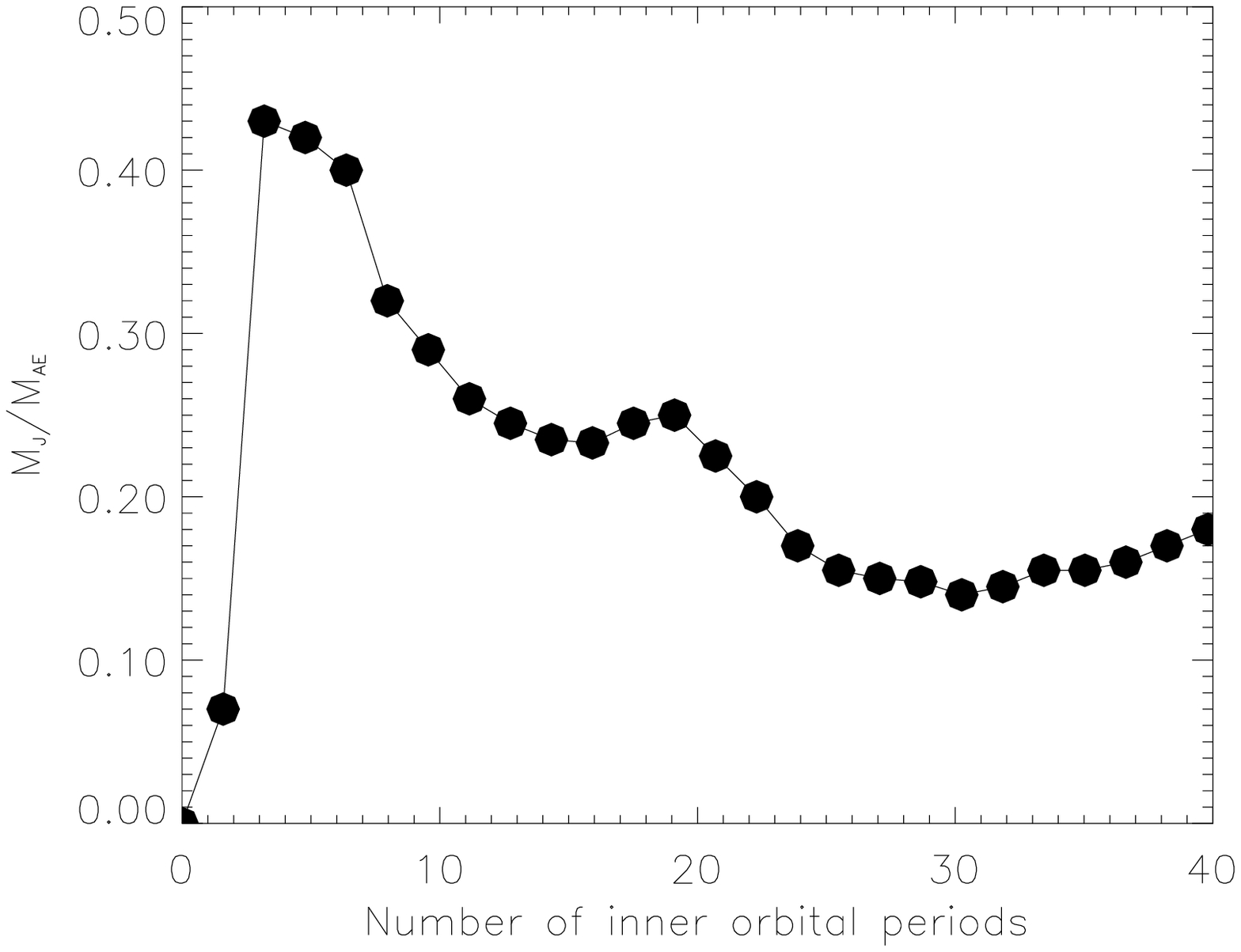}
\caption{Time evolution of the ejection rate of the structure
$\dot{M}_j$. After the jet is launched, the ejection rate slowly
decreases and then reaches a plateau. This indicates that the ejection
mechanism is robust
once the jet is formed. The typical amount of matter
ejected from the accretion disk is roughly $\dot{M}_j/\dot{M}_{ae}\sim
15\%$ in this phase.\label{Ejec}}
\end{figure} 

 Another way to determine if the obtained
emission of matter can be qualified as a robust mechanism, is by
measuring the ejection rate of matter from the disk. To that end,
we measure the mass flux through the surface of the disk, namely the
surface where the radial velocity is vanishing. Since we impose the
external accretion rate in the disk by means of our `source' boundary condition
(see previous paragraph), we can then obtain the ratio of the ejection rate
$\dot{M}_j$ to the external accretion rate $\dot{M}_{ae}$. This ratio
evolves in time as shown in Fig.~\ref{Ejec}. This plot shows
two to three distinct phases. The first one marks the beginning of the jet
launching from the disk where the ejection rate is increasing. The second
phase corresponds to a relaxation of the system. Finally,
the ejection rate reaches a `plateau' at the end of our simulation, from
about 25 rotation periods onwards. This
plateau indicates that the ejection mechanism is robust and jet material
is constantly launched from the disk.
We did not perform a simulation to even larger integration times since
these simulations are very time-consuming. 

\begin{figure}[t]
\plotone{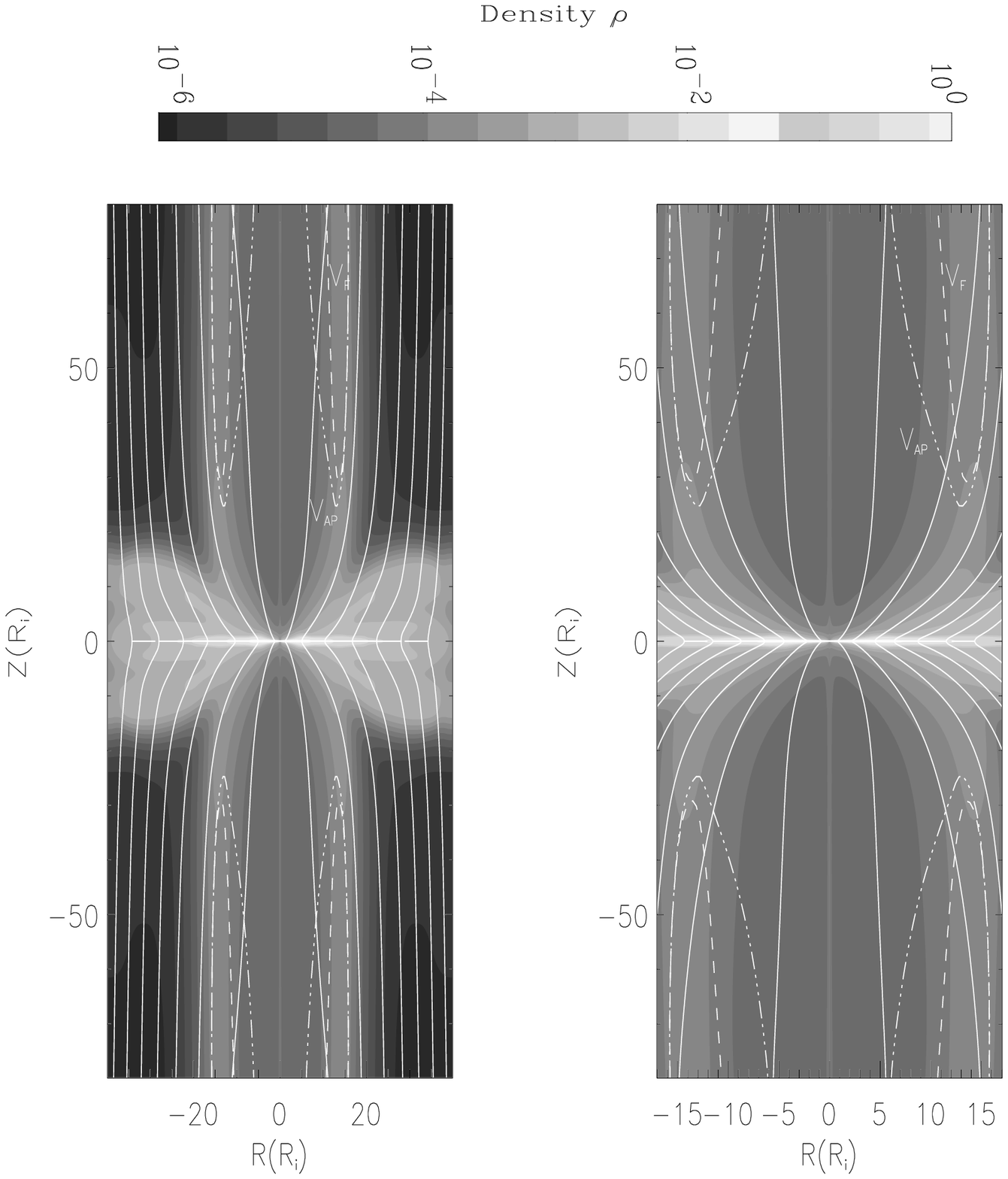}
\caption{Snapshot of the accretion-ejection 
structure at $t=19$. On the left the complete
computational domain is represented by setting density levels using
greyscales while \pol magnetic field lines and critical surfaces are
displayed in solid and dot-dashed lines, respectively.
The right panel is a zoom of the jet launching region ($R\leq 20$).\label{Snap2}}
\end{figure}

In the next subsection,
we give a detailed description of a 
snapshot at $t=19$ of our simulation where the system has settled
on a {near-stationary} state. This snapshot is shown in Fig.~\ref{Snap2} where 
the density is displayed by greyscales, the \pol magnetic field lines by solid
lines. Also indicated are the Alfv\'en ($V_{A,p}$) and fast-magnetosonic
($V_F$) surfaces, as 
represented by
dotted-dashed lines. The considered velocities are defined as
\be 
V_{A,p}^2=\frac{B^2_p}{\rho} \ ; \ V_F^2 = \frac{1}{2}\left\{C_S^2+V_A^2 + \sqrt{(C_S^2+V_A^2)^2-4C_S^2V_{A,p}^2}\right\}
\label{caract}
\ee
\noindent where $V_A$ is the total Alfv\'en speed
$V^2_A=(B_p^2+B^2_{\theta})/\rho$. We can see in this
snapshot that most of the matter of the jet has been accelerated to
super-fast flow speeds
at the top of the computational domain. The collimation is almost complete
at the top of the simulation box. Note also the distinct hollow jet structure,
with the fastest moving material situated in a dense cylindrical shell at
a radius $R\simeq 13$.   
\subsection{Accretion-ejection mechanism}           
\begin{figure}
\plotone{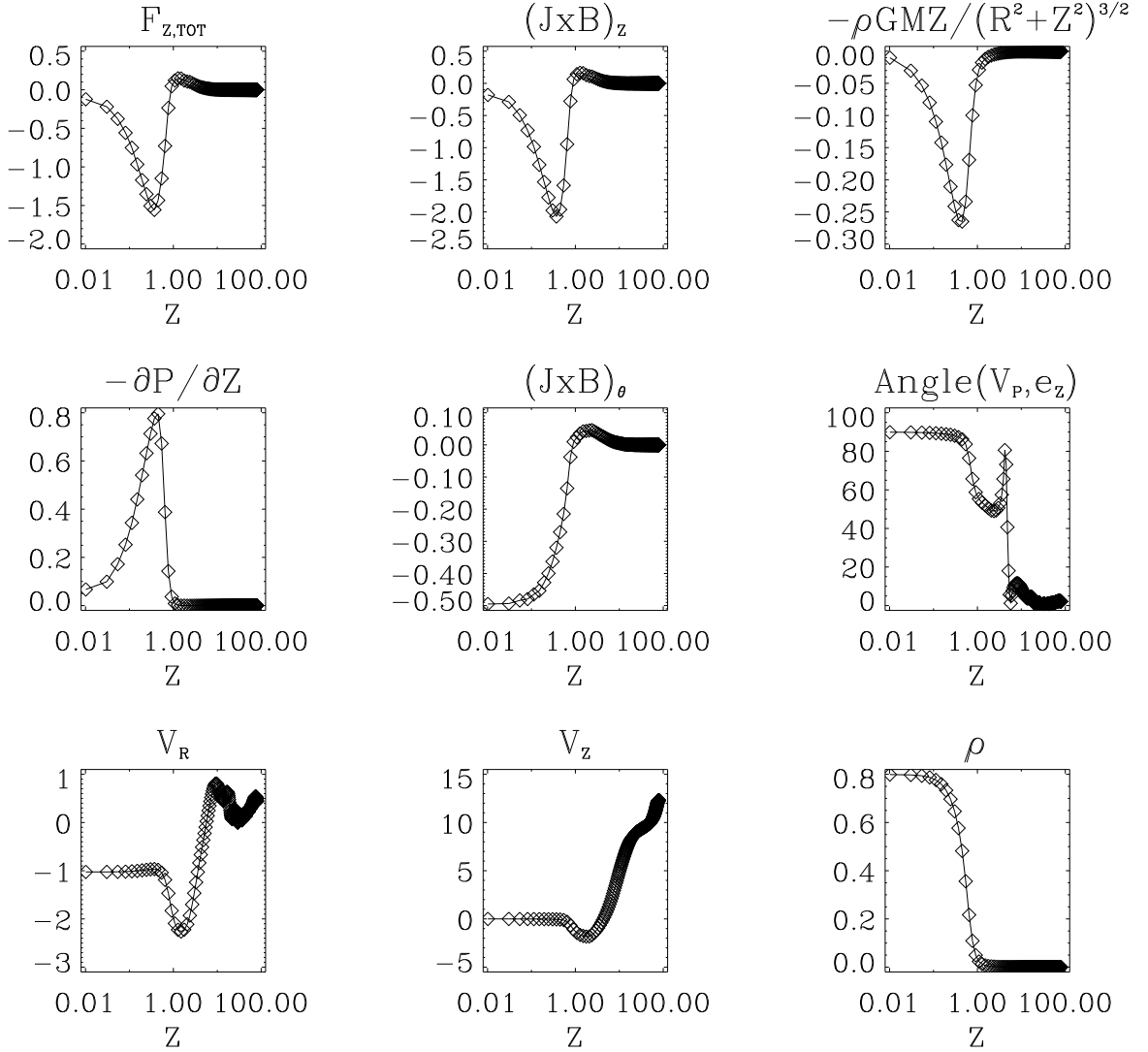}
\caption{Vertical profiles of several quantities at a given radius $R=4.3$
and at $t=19$. From left to right, top to bottom:
total vertical force balance, vertical component of the Lorentz force,
vertical gravity component, vertical pressure gradient force, magnetic
torque, angle between the poloidal velocity and the vertical, both poloidal
flow components, and the density structure.
On the corresponding streamlines plot shown in Fig.~\ref{AE2},
two regimes are occuring: the first one shows streamlines reaching the inner
radius in a pure accretion motion while the second one are streamlines which
start to accrete but then turn back into an ejection motion. This can be
explained by looking at the vertical cuts shown: the \tor component
of the magnetic force $({\bf J}\times\BB)_\theta$ 
is negative inside the disk ($Z\leq 1$) and changes
its sign near the disk surface. A \pol acceleration takes place once the
sign reversal is achieved as we can see in $V_R$ and $V_Z$. These
two components of \pol velocity stop decreasing as soon as 
$({\bf J}\times\BB)_{\theta}$ turns positive. \label{AE}}
\end{figure}
\begin{figure}[t]
\plotone{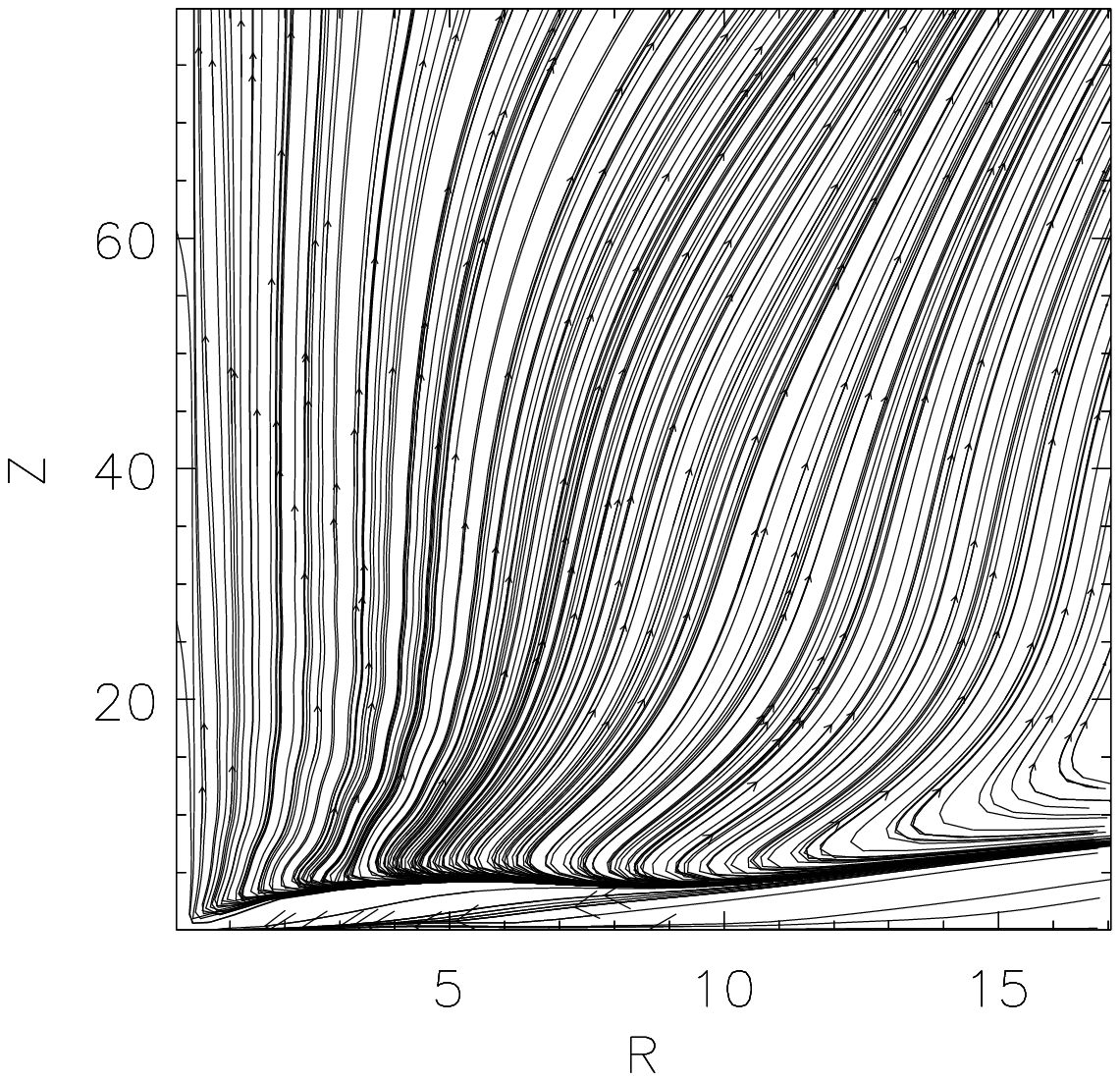}
\caption{Polo\"{\i}dal streamlines of the accretion-ejection structure at
t=19. Two regimes can be distinguished: the first one shows streamlines reaching the
inner radius in a pure accretion motion while the second one contains 
streamlines
which start to accrete but then turn into an ejection motion. This
can be explained by Fig.~\ref{AE} which displays the accretion-ejection
mechanism.~\label{AE2}}
\end{figure}
In this subsection, we give a complete description of the accretion-ejection
mechanism that forms spontaneously in our simulation. The 
mechanism is driven by the
magnetic field which plays a double role. Its first role is to brake the
matter in the disk and transfer angular momentum of matter to allow accretion. 
The magnetic torque (\tor
component of the magnetic force ${\bf J}\times\BB$) must therefore be negative
in the disk. The second role of the magnetic field is to accelerate matter
in the jet by providing a positive projected magnetic force along the
streamlines. Such a positive projected force component 
is achieved if the magnetic
torque changes its sign. Indeed, 
it is obvious that 
\be
({\bf J}\times\BB).\BB_p = -({\bf J}\times\BB).\BB_{\theta}. 
\label{AE1}
\ee
\noindent Hence, if we assume that the \pol streamline is almost (or
completely, in a stationary framework) parallel to the \pol magnetic field
line, we have in the jet $\vv_p\sim \alpha\BB_p$.  Since $B_\theta$ is
negative,  the magnetic torque must be positive in the jet for accelerating
matter upwards along the magnetic field direction.  A positive torque spins
up jet material, and this azimuthal acceleration of the plasma  increases
the angular momentum of matter which leads to a dominant centrifugal force 
that will tend to widen the magnetic surfaces.  Note that ideal MHD
reasoning makes sense since the resistivity is equal to zero in the jet
region. 

In Fig.~\ref{AE} and Fig.~\ref{AE2}, the entire mechanism is well
illustrated by showing both the streamlines of the flow in the jet
launching region and several characteristic quantities along a vertical cut
at a fixed radial distance of $R=4.3$. The magnetic torque ${\bf J}_p
\times\BB_p$ is seen to reverse its sign at about $Z=1$ and provokes a
reversal of the \pol velocity vector. The accretion-ejection needs one more
condition to work. Indeed the change of sign of the magnetic torque is a
necessary condition in order to ensure that the jet will receive energy for
accelerating the matter via a MHD Poynting flux. But another necessary
condition is that the accretion disk must provide mass for the jet. The
mass flux, and more precisely a vertical mass flux can only be achieved by
a vertical force balance in the accretion disk which becomes positive
(upwards) at the disk surface.  At the same time, the vertical equilibrium
in the interior  of the accretion disk must be such that the total vertical
force is negative and thereby keeps the main part of the plasma inside the
accretion disk. This delicate force balance ensures that only a small
fraction of the accretion disk matter escapes. If we look at Fig.~\ref{AE},
we see that the vertical equilibrium configuration is occuring in precisely
the manner mentioned here.  Plotted are all vertical forces acting on the
disk, so we can identify which ones pinch the disk and which ones lift the
matter up. The density profile of our accretion disk (dense near the
equatorial plane and decreasing vertically) leads to a positive thermal
pressure gradient that will lift the matter. On the contrary, the
gravitational force pinches the disk as well as the magnetic pressure.
Indeed, due to the shape of the magnetic surfaces, a growth of the radial
and azimuthal component of the magnetic field occurs.  This bipolar
configuration leads automatically to a magnetic pinching of the accretion
disk. As seen in Fig.~\ref{AE} the main competition in the vertical balance
is between thermal and magnetic pressure gradients, revealing that the
thermal pressure must not be smaller that the magnetic one. Moreover if the
magnetic pressure (i.e. magnetic field) is too small, the first condition
mentioned at the beginning of this paragraph will not be fulfilled. The
best configuration for emitting a fast jet of matter from an accretion disk
seems then to be an accretion disk in equipartition between thermal
pressure and magnetic pressure, as already noticed by \citet{Ferr95}. 

\begin{figure}
\plotone{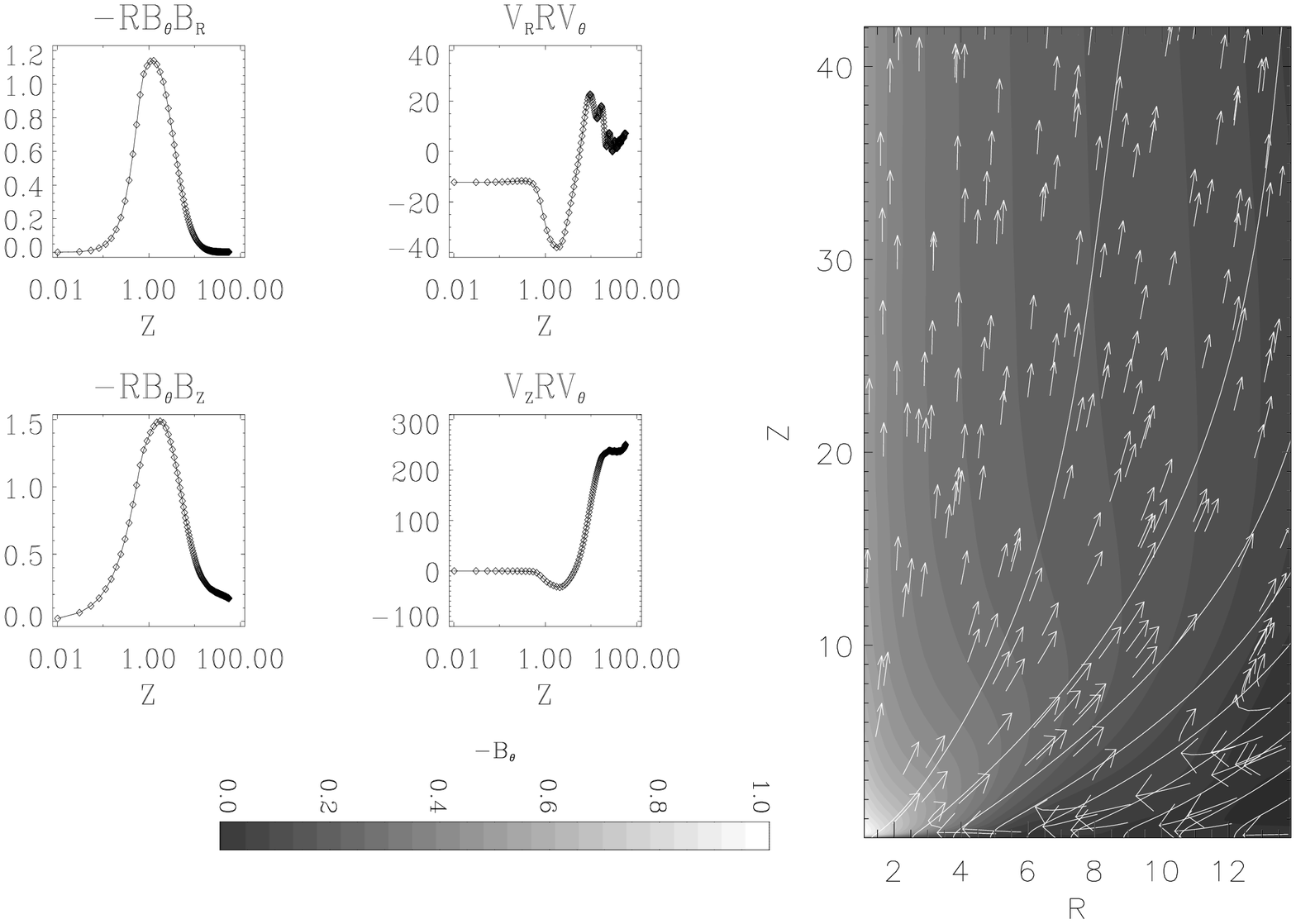}
\caption{ Left: magnetic and matter angular momentum contributions to the
total angular momentum flux. Right: vector map of the total angular
momentum flux ${\bf F}_{AM}=\rho V_{\theta}R{\bf V}_p -
RB_{\theta}\BB_p$ (arrows), \pol magnetic surfaces (solid lines) and
amplitude of \tor magnetic field (greyscale). The left panels represent
the different contributions to the total angular momentum flux in the
radial and vertical directions. Note the
extraction of angular momentum by the generation of $RB_{\theta}\BB_p$ in
the disk and secondly an increase of the specific angular momentum of jet
matter due to a magneto-centrifugal acceleration. On the right
panel, we can see that the infalling radial flux is evacuated in the jet
along the magnetic surfaces and that the turning point occurs where
$B_{\theta}$ reaches its maximum value, namely near the disk
surface. \label{f11}}
\end{figure}

From an energetic point of view, the accretion-ejection mechanism decribes the
transfer of the angular momentum of the disk into the jet. Indeed,
the \tor component of the momentum equation Eq.~(\ref{mhd2})
expresses angular momentum conservation as
\be 
\frac{\p \rho V_{\theta}R}{\p t} + {\bf \nabla}\cdot\left(\rho V_{\theta}R{\bf
V}_p -RB_{\theta}\BB_p\right)= 0 \ .
\label{AEbis}
\ee
\noindent The action of the magnetic field provides a way to extract the
angular momentum of disk matter and thus enable accretion inside the
disk. 
At the opposite, in the jet, the angular momentum stored in the magnetic field
(via the generation of a \tor component $B_{\theta}$), can be used to
accelerate matter in order to power a magneto-centrifugal jet. On
Fig.~\ref{f11}, we display both a vertical cut and a full map of the
total angular momentum flux ${\bf F}_{AM}=\rho V_{\theta}R{\bf V}_p -
RB_{\theta}\BB_p$. The left panels represent the matter and magnetic contributions to this
flux in the radial and vertical directions. The dominant patterns are first,
the storing of angular momentum by the generation of $RB_{\theta}\BB_p$
in the disk and secondly an increase of the specific angular momentum of
matter in the jet due to a magneto-centrifugal acceleration. On the
right panel, 
the orientation of ${\bf F}_{AM}$ is displayed and we can see that the infalling
radial flux in the disk is transferred to the jet along the magnetic
surfaces. Note that the acceleration of matter decreases as soon as
the angular momentum stored in the magnetic field reaches zero.    
\begin{figure}
\plotone{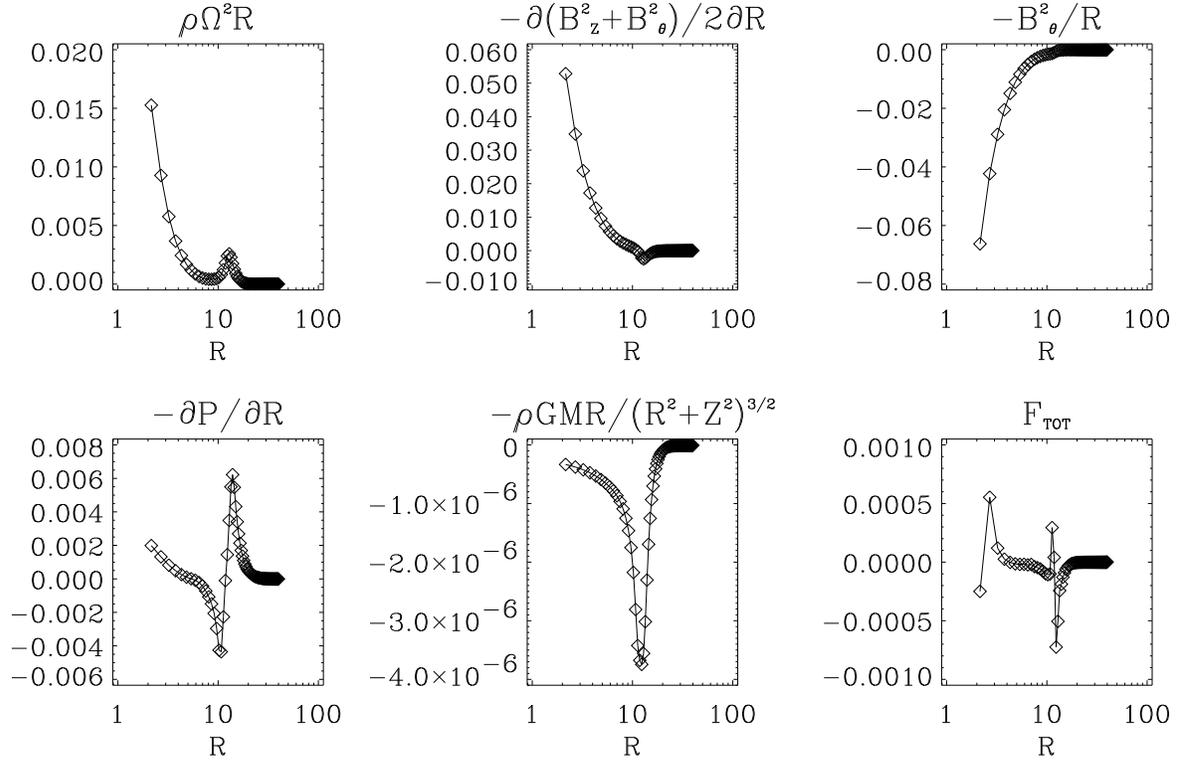}
\caption{Radial cut of the jet structure 
representing radial forces acting on the
plasma at $Z=40$. Two classes of forces are acting on the plasma: the ones
that try to widen the flow and those that try to collimate. The
first class is composed of the centrifugal force and the total pressure, while
the other one contains gravity and the magnetic tension due to the
presence of a \tor magnetic field. This figure illustrates that
only the presence of $B_{\theta}$ is able to counteract other forces in the
jet region where gravity is almost negligible. Note that we do not involve
any external mechanism to ensure the collimation of the flow. \label{Colli}}
\end{figure} 
\begin{figure}
\plotone{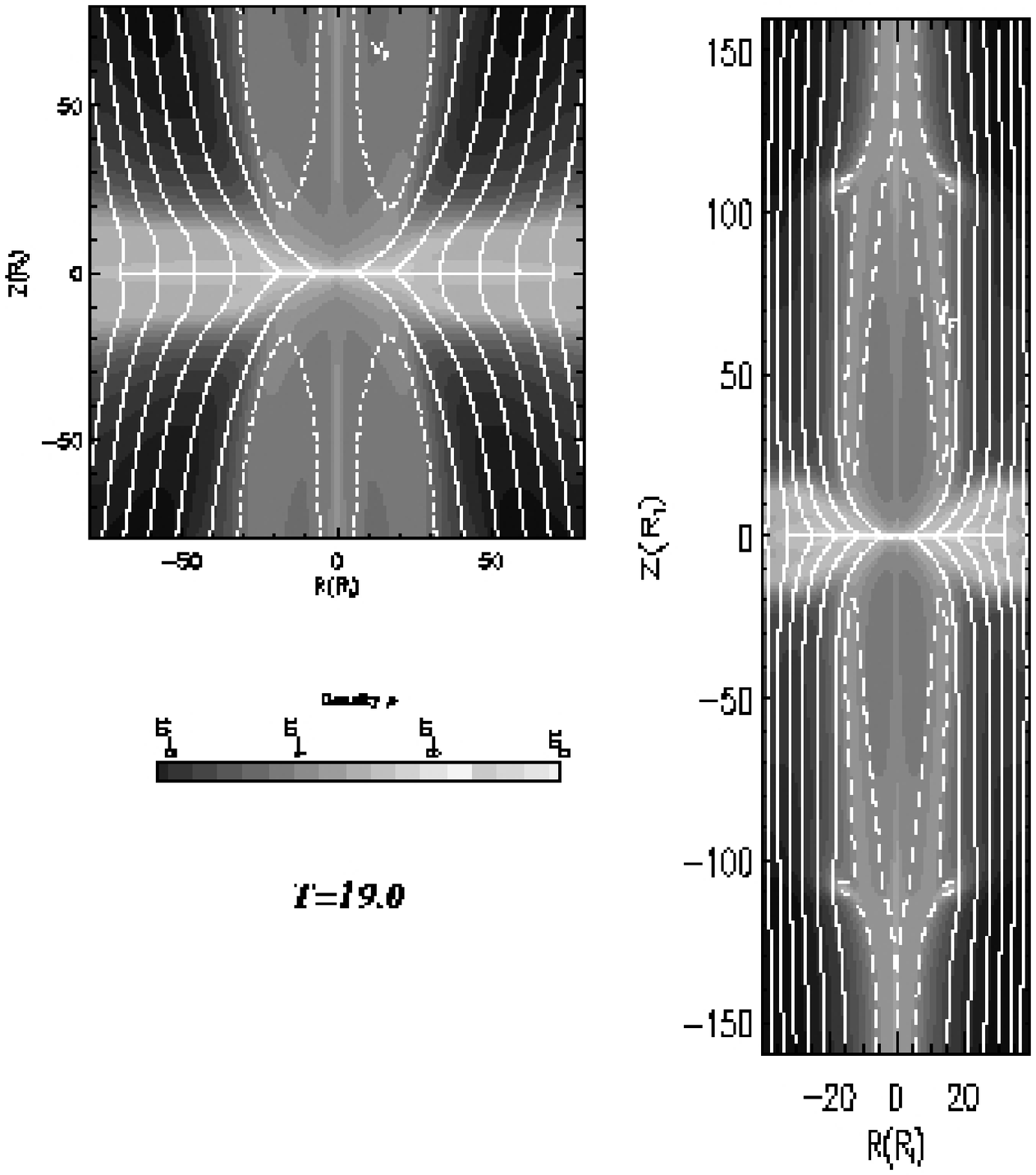}
\caption{Snapshots of two equivalent MHD simulations on extended
domains using the same initial conditions as
the reference simulation of Fig.~\ref{Snap} and at the same physical time
$t=19$, to be compared with Fig.~\ref{Snap2}. 
The influence of the boundary conditions is marginal, as all
simulations reach very similar super-fast collimated outflow
launched from the inner disk region alone.\label{Box}}
\end{figure}

It is important to note that after the acceleration has taken place, the
collimation of the flow is quite good since the maximal value of the angle
between the vertical direction and the velocity in the highest regions of
our box is never bigger than $1.5^{\circ}$ (see Fig.~\ref{AE}).  The
collimation of the flow is due to a mechanism internal to the jet structure
being formed.  This is demonstrated in Fig.~\ref{Colli}, showing all the
radial forces existing in the jet at a horizontal cut at $Z=40$.  These are
the centrigugal force, the thermal and magnetic pressure gradients, gravity
and magnetic tension.  Clearly, the force assuring the collimation of the
flow is the magnetic tension or ``hoop'' stress
($B^2_{\theta}/R$). Contrary to all other forces (except gravity which can
be neglected so far from the disk), it is the only force that is always
directed towards the symmetry axis. Looking at the sum of all forces in the
radial direction indicates that the plasma is not yet in a full stationary
state but has come quite close to achieving a radial equilibrium balance.
However, we can not exclude the possibility that eventually some
hydrodynamic or magnetic instabilities develop somewhere in the
structure \citep{Kim00}. This  should be verified by performing the same simulation in
three dimensions.

\subsection{Influence of open boundaries}  

\begin{figure}[t]
\plotone{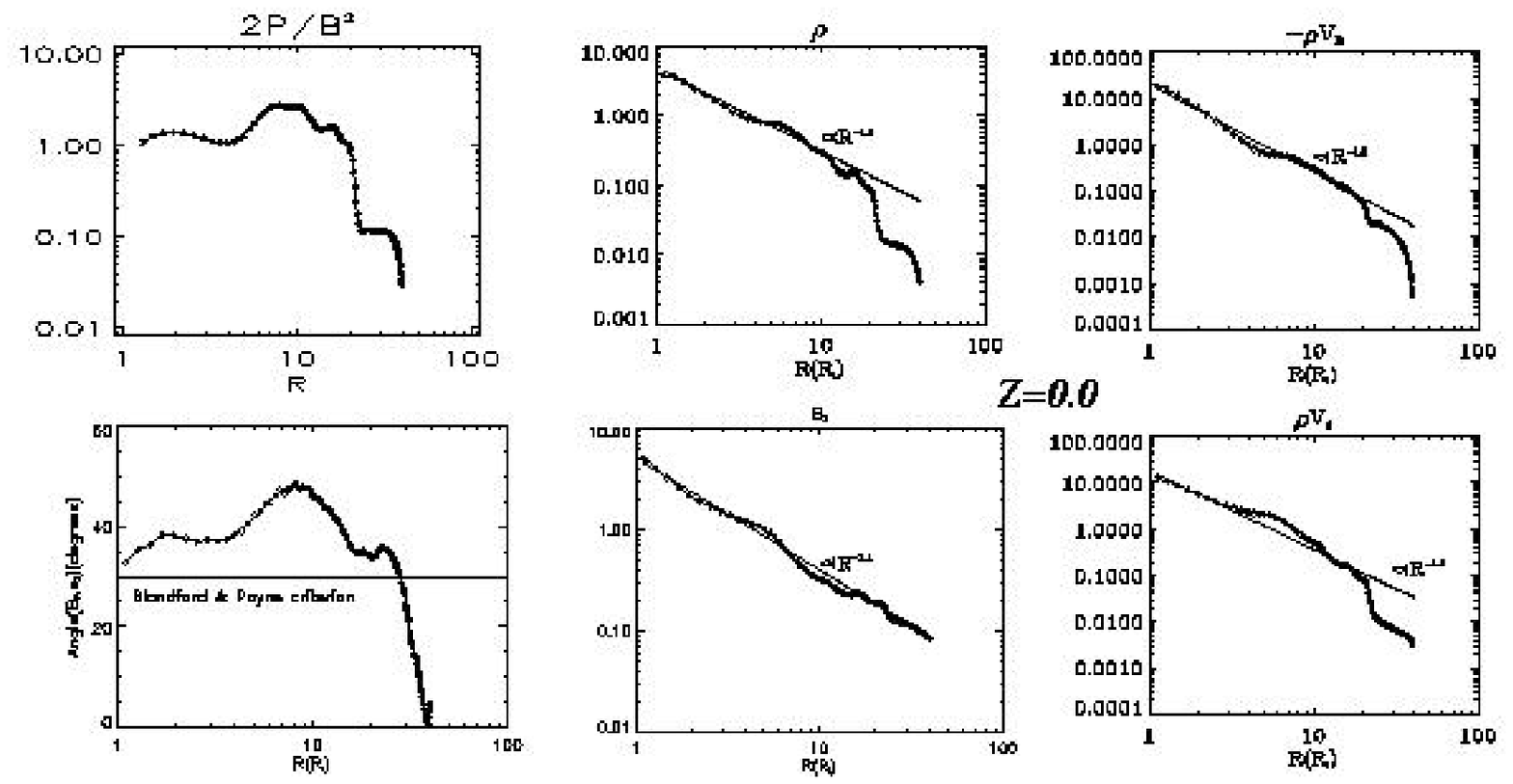}
\caption{Ratio of thermal pressure to magnetic pressure, radial profiles
of physical quantities ($\rho,-\rho V_R, B_z, \rho V_{\theta}$) along the
equatorial plane ($Z=0$), and angle between $\BB_p$ and ${\bf e}_z$ at the disk surface (where the
radial velocity is vanishing). The $\beta$ 
ratio explains why the inner part of the disk launches a jet
while the outer part does not: only the inner part is in
equipartition,
a necessary condition ensuring that the
vertical disk equilibrium can supply matter for the jet
formation. Moreover, as shown by the bottom left plot, only the inner
region has a suitable magnetic configuration for matter acceleration in the
jet. The behaviour of quantities along the equatorial plane is close
to radial power-law with indices modified in comparison to the
initial conditions. The power-law indices are in relatively good agreement
with self-similar MAES models from \cite{Cass00a}.\label{ADs}}
\end{figure} 
Modeling a magnetized, resistive accretion disk and its associated ideal
MHD trans-Alfv\'enic jet numerically requires to control the effect of, in
particular, the open boundaries bordering the computational domain. Indeed,
as already shown by \cite{Usty95}, if not modeled appropriately, they can
influence the dynamics of the system, e.g. by unwanted reflections. To
check the potential effect of our boundary conditions on the simulated
flows, we repeated all our simulations using different sizes for the
computational domain. While our reference simulation uses a $[0-40]\times
[0-80]$ rectangular box, we performed the same calculation with rectangular
boxes covering $[0-80]\times [0-80]$ and $[0-40]\times [0-160]$, with
resolutions of $204\times 104$ and $104\times 404$, respectively.  The jet
launching was found to occur in every simulation and, most importantly,
remains restricted to roughly the same part of the physical domain (within
$1-20$ radius range), as shown in Fig.~\ref{Box}.  Nevertheless, small
differences do exist between these three simulations. The radial extension
of the computational domain seems to have an influence on the exact radial
size of the jet launching region. This indicates a marginal influence from
the open  boundary at the right side of the box. Repeating the simulations
while gradually extending the radial box size, a systematic larger jet
launching radial range is obtained. The effect is probably also due to
differences in effective local resolution, but can be observed by comparing
Fig.~\ref{Snap2}  with the left panel of Fig.~\ref{Box}.  The simulation
which doubled the vertical extent (right panel of Fig.~\ref{Box}) is in
very good agreement with the reference solution. Hence the top open
boundary has no influence on the obtained dynamics.

\subsection{Accretion disk structure}

The radial accretion disk structure largely determines its stability to
MHD perturbations~\citep{apjl}.
Up to the present work, the only model able to include both a
resistive accretion disk and a super-Alfv\'enic jet without neglecting
any dynamical quantities in the equilibrium of the structure
was a self-similar model used by \citet{Ferr97} and \citet{Cass00a}. This
model assumed  
that, to simplify the analysis, all physical quantities $A$ 
appearing in the set of MHD equations can be written
as 
\be
A(R,Z)= A_e\left(\frac{R}{R_e}\right)^{\alpha_A}f_A\left(\frac{Z}{R}\right),
\label{Diskstru1}
\ee
\noindent where $\alpha_A$ is a coefficient imposed by the smooth crossing
of the Alfv\'enic surface. Hence, each quantity was represented by a
power-law dependence on radius $R$ and a separate polar angle variation
$f_A$.  Since the initial conditions of our simulation are derived from a
self-similar configuration, it is worthwhile  to confront this with the
obtained final radial structure of our accretion disk. In Fig.~\ref{ADs},
we display the radial profiles of several physical quantities at the disk
midplane $Z=0$. We clearly see two different regions. The first one which
reaches up to $R\leq 20$ corresponds to the jet launching region  and
features a radial behaviour of all quantities close to power-law. As
compared to their $t=0$ values, these indices are modified.  Indeed, in our
initial conditions, the indices for density $\rho$, momentum $\rho\vv$ and
magnetic field $B_z$ were $-3/2, -2$ and $-5/4$. These initial values
correspond, in the case of self-similar MAES, to a magnetized accretion
disk where no ejection occurs.  In the final disk structure shown in
Fig.~\ref{ADs}, the indices have become $\alpha_{\rho} > -3/2$,
$\alpha_{\rho v} > -2$ and $\alpha_{B} > -5/4$.  Such a systematic
modification to larger values is consistent with self-similar MAES models
that allow ejection.  Indeed, the presence of a jet makes the effective
local accretion rate a decreasing function for smaller radii and this
modifies the power-law radial indices of the quantities.  Our simulation
shows that the final disk structure displays a second region where the
quantities, except for the magnetic field, do not have a clear power-law
behaviour.  This explains why this outer disk region does not launch a
jet. Indeed, the parameter $\beta=2P/B^2$ which was set to a constant value
of $1.67$ at $t=0$, has adjusted to a very low value outside the jet
launching region.  As explained before, this prevents the vertical force
balance in the disk to allow matter escaping from the disk and providing a
mass source for the jet. Moreover, the angle of the \pol magnetic field
with respect to the vertical direction measured at the disk surface (where
resistivity is vanishing), is suitable for jet launching only in the inner
region. Indeed \citet{Blan82} has shown that any cold steady-state jet must
have, bent \pol magnetic surfaces with an angle larger than thirty degrees,
in order to achieve an acceleration by the action of the magnetic
field. In Fig.~\ref{ADs}, on the left bottom panel, we show the angle
between the magnetic surface and the vertical direction measured at the
disk surface (typically where $V_R$ changes its sign). It is clear from
this figure that only the inner part of the structure is able to launch a
jet. Nevertheless, the suitable bending of the magnetic surfaces is ranging
from $R=1$ to $R\simeq 30$ in contrast with the effective jet launching
region that ranges from $1$ to $\sim 15$. This is a good illustration that
both jet launching conditions must be fulfilled: in the region between $16$
to $30$, the vertical balance of the disk is not suitable for a positive
mass flux). 
    
\section{Concluding remarks and outlook}

We present in this work the first MHD simulations of a magnetized, resistive
accretion disk launching an ideal trans-Alfv\'enic jet. The simulations are
performed in a 2.5  dimensional, time-dependent and polytropic framework.
Our MHD simulations involve a magnetic resistivity that is only triggered
within the accretion disk. The shape of the resistivity coefficient is
derived from \citet{Shak73} since we express it as $\eta=\alpha_m
V_A|_{Z=0} H\exp(-2Z^2/H^2)$ where $V_A$ stands
for Alfv\'en velocity, $H$ disk scale-height and $\alpha_m$ is a dimensionless parameter
controlling the amplitude of the resistivity, smaller than unity). The temporal behaviour of
our simulations display the launching of an outflow that, while
propagating, becomes collimated. Note that the jet reaches
super-fastmagnetosonic velocities within our computational domain.  The
long term robustness of the structure is demonstrated by several means.  In
particular, we check the angle between \pol velocity and \pol magnetic
field (which is zero in a pure ideal MHD stationary configuration) and show
that, after the outflow has been launched, this angle reaches maximum
values in the jet typically smaller than three degrees. Moreover, the
ejection rate (mass flux through the disk surface) of the disk, once
initiated, remains near-constant over  several tens of the accretion disk
dynamical timescale.  These diagnostics characterize a robust system that
has reached a  quasi-stationary state. 

We also present in this paper, by a detailed scrutiny of one snapshot of our
simulation in the near-stationary phase, 
a complete illustration of the so-called `accretion-ejection'
mechanism that is responsible for the jet launching. The key point of this
model is that disk resistivity enables the structure to reach a
near-static 
equilibrium where accreted matter inside the disk can pass through the
\pol magnetic surface. Such resistivity would mimic the effects of
turbulence occuring inside the disk, provoked by MHD instabilities
affecting equipartion accretion disks \citep{apjl}. Moreover, we confirm
by self-consistent numerical simulations
that the two necessary conditions to achieve such a jet creation are (1) a
vertical disk equilibrium which allows some disk matter streamlines to reach
the disk surface (typically an equipartition disk with $B^2/P\sim 1$,
\citet{Ferr95}) and (2) a turbulence configuration that obeys a
simple relation Eq.~(\ref{init6}) already presented in \citet{Cass00a}. 
In our simulations, since we do not take viscosity into account, the
angular momentum of disk matter is transferred by the action
of a magnetic torque acting on the disk plasma. We highlight the fact that
the sign reversal of the magnetic torque at the disk surface, obtained in
self-similar analytic models if the
turbulence follows the relation \ref{init6}, is the core of the
magneto-centrifugal effect responsible for jet
acceleration. By displaying the radial force balance in the jet, we
illustrate the role of the \tor component of the magnetic field which
collimates the outflow by means of magnetic tension (also called
``hoop'' stress, see \citet{Heyv89}). 

Forthcoming work will focus on two major points that were not taken into
account here. Indeed, future work should be devoted to the implementation
of a viscous torque inside the disk. This approach will validate the
knowledge on complex angular momentum transport by both magnetic and
viscous torques \citep{Cass00a,Ogil01} combined with jet launching. The
second point is the relevance of the energy equation in this kind of
system.  Specifically, it has been demonstrated in a self-similar framework
that the existence of a hot corona located at the disk surface is relevant
to the astrophysical outflow \citep{Cass00b,Garc01}. So the use of a full
energy equation is desirable.  The simulations presented here consider a
time-dependent resistivity (through the value of the Alfv\'en speed in the
disk). This could be of great interest for scenarios looking at
time-dependent disk models launching jets, in particular for microquasars
\citep{Tagg99,Naya00}. Indeed, a suitable prescription of the transport
coefficients could be able to produce structures where a periodic jet
launching occurs.

\acknowledgments
The authors are very grateful to Hans Goedbloed for fruitful discussions
and for his comments on the manuscript.
This work was done under Euratom-FOM Association Agreement with
financial support from NWO, Euratom, and the European Community's Human
Potential Programme under contract HPRN-CT-2000-00153, PLATON, also
acknowledged by F.C. NCF is acknowledged for providing computing facilities.

\end{document}